\documentclass[]{jfm}

\usepackage{graphicx}
\usepackage{newtxtext}
\usepackage{newtxmath}
\usepackage{natbib}
\usepackage{hyperref}
\usepackage{soul}
\hypersetup{
    colorlinks = true,
    urlcolor   = blue,
    citecolor  = black,
}

\newcommand{\RomanNumeralCaps}[1]

\shorttitle{Nonlinear fluid damping of pitching wings}
\shortauthor{Y. Zhu, V. Mathai and K. Breuer}

\title{\vspace{-1.8cm}Nonlinear fluid damping of elastically mounted pitching wings in quiescent water}

\author{Yuanhang Zhu \aff{1}
  \corresp{\email{yuanhang\_zhu@brown.edu}},
  Varghese Mathai \aff{2}
 \and Kenneth Breuer \aff{1}
 }

\affiliation{\aff{1}
Center for Fluid Mechanics, School of Engineering, Brown University, Providence, RI 02912, USA
\aff{2}
Department of Physics, University of Massachusetts, Amherst, MA 01003, USA}

\begin{document}
\maketitle

\begin{abstract}
We experimentally study the nonlinear fluid damping of a rigid but elastically mounted pitching wing in the absence of a freestream flow. The dynamics of the elastic mount are simulated using a cyber-physical system. We perturb the wing and measure the fluid damping coefficient from damped oscillations over a large range of pitching frequencies, pitching amplitudes, pivot locations and sweep angles. A universal fluid damping scaling is proposed to incorporate all these parameters. Flow fields obtained using particle image velocimetry are analyzed to explain the nonlinear behaviors of the fluid damping.

\end{abstract}
\rule{0.95\textwidth}{0.5pt}
\vspace{10pt}


\section{Introduction}\label{sec.intro}

The interaction between elastically mounted pitching wings and unsteady flows is central to many applications. With a free-stream flow, this interaction can lead to self-sustained, flow-induced oscillations, which have been studied for understanding classic aeroelastic behaviour \citep{dowell1989modern,dugundji2008some}, as well as in developing oscillating foil energy harvesting devices \citep{xiao2014review,young2014review}. Without a free stream, but with prescribed heaving or flapping (i.e. hovering), the passive flow-induced pitching motions are used in modelling the thrust generation and maneuvering in animal flight \citep{wang2005dissecting,bergou2007passive,shinde2013jet,kang2014analytical,beatus2015wing}.

One of the critical parameters that govern the flow-structure interactions of passively pitching wings is the fluid damping. According to the semi-empirical Morison equation \citep{morison1950force}, the total fluid force exerted on a wing submerged in unsteady viscous fluid can be divided into two parts -- the force associated with fluid inertia (i.e. the added mass force), which is in phase with acceleration \citep{brennen1982review,corkery2019quantification}, and the force induced by vortices in the flow (i.e. the fluid damping force), which is in phase with velocity \citep{shih1971drag,kang2014analytical,su2019resonant}. While the structural damping force is typically proportional to velocity because of the constant structural damping coefficient, the fluid damping force is expected to scale quadratically with velocity \citep{morison1950force,keulegan1958forces}, and due to this nonlinearity, the fluid damping coefficient is usually obtained empirically as a function of the reduced frequency, the Reynolds number, the oscillation amplitude, etc \citep{shih1971drag}. For pitching flexible wings \citep{alben2008optimal} and heaving membrane wings \citep{tzezana2019thrust}, the fluid damping coefficient is found to scale inversely with the oscillation frequency.

For elastically mounted pitching wings with a free stream, the interplay between the fluid damping and the structural damping governs the flow-induced oscillation. By mapping out the cycle-averaged energy transfer between the elastic system and the ambient fluid using prescribed kinematics, \citet{menon2019flow} and \citet{zhu2020nonlinear} showed that the energy injected by the negative fluid damping must be equal to the energy dissipated by the positive structural damping in order for the flow-induced oscillations to sustain. In other words, the total damping of the system must be zero \citep{dugundji2008some}. The negative fluid damping arises primarily from the formation and shedding of dynamic stall vortices \citep{mccroskey1982unsteady,corke2015dynamic}. In the absence of a free stream, however, the fluid damping becomes positive and counteracts the pitching motion because of the drag effect. With both the fluid damping and the structural damping being positive, any perturbations to the system will be damped out. However, little is known about how the fluid damping shapes the damped oscillations, and understanding this is of critical importance for understanding the fluid-structure interactions of elastically mounted pitching wings under external perturbations such as gusts.

In the present study, we use laboratory experiments to characterise the fluid damping of elastically mounted pitching wings in quiescent water, with the elastic mount simulated using a cyber-physical system (\S \ref{sec.setup}). We perform `ring down' experiments to extract the fluid damping (\S \ref{sec.ringdown}). The effects of many parameters are investigated, including the effects of the pitching frequency, the pitching amplitude, the pivot location and the sweep angle (\S \ref{sec.freq}). We propose a universal fluid damping scaling to incorporate these parameters (\S \ref{sec.scaling}), and correlate the nonlinear behaviour of the fluid damping with the dynamics of the vortical structures measured using particle image velocimetry (\S \ref{sec.flow_dynamics}). Finally, the key findings are summarised in \S \ref{sec.conclusion}.

\section{Experimental set-up}\label{sec.setup}

\begin{figure}
\centering
\includegraphics[width=.75\textwidth]{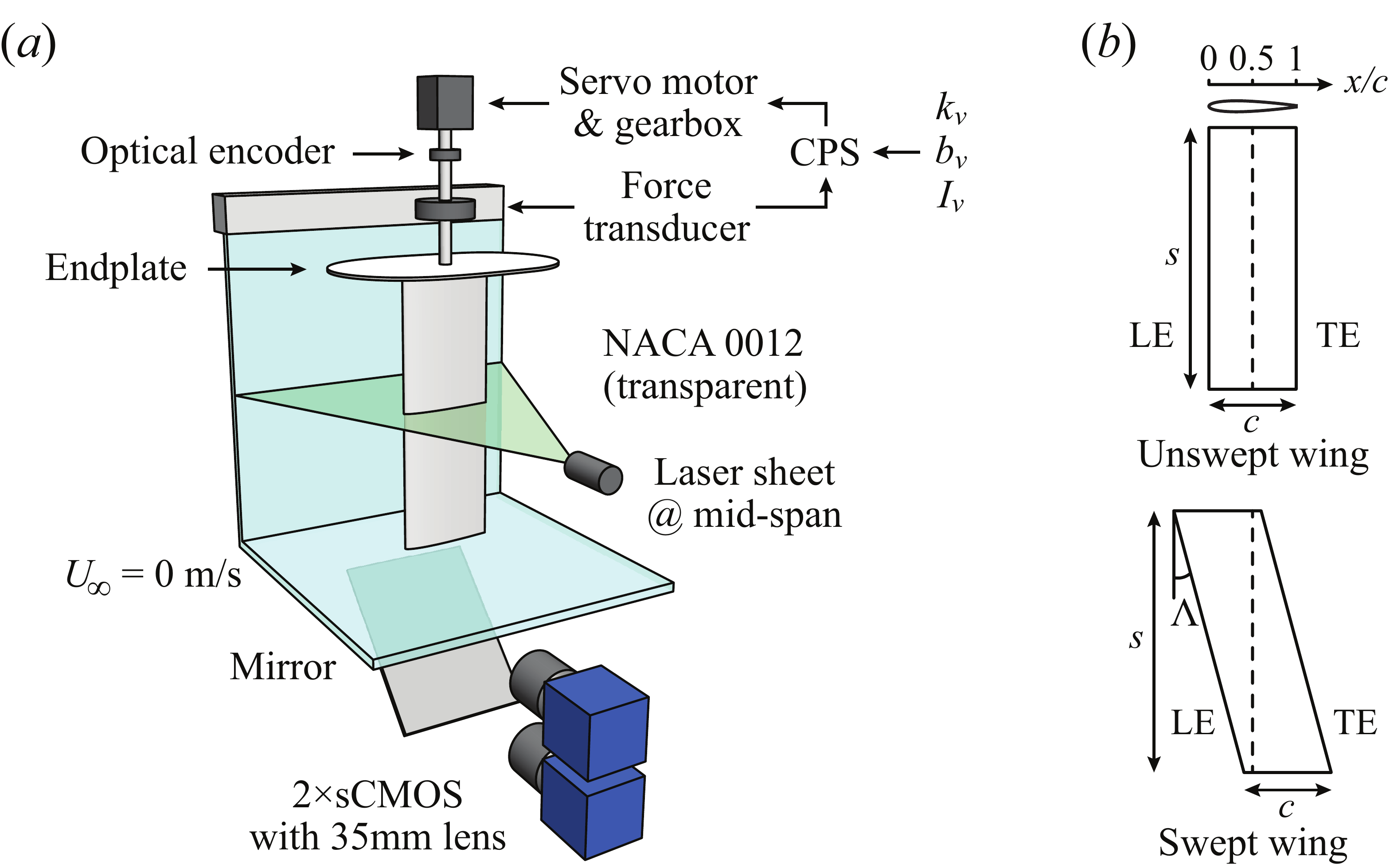}
\caption{(\emph{a}) A schematic of the experimental set-up. The structural dynamics of the wing is simulated by a cyber-physical system (CPS). (\emph{b}) Sketches of unswept and swept wings. The leading edge (LE) and the trailing edge (TE) are parallel. Dashed lines represent the pivot axis.}
\label{fig.setup}
\end{figure}

Figure \ref{fig.setup}(\emph{a}) shows a schematic of the experimental set-up. We conduct all the experiments in the Brown University free-surface water tunnel (test section $\mathrm{width} \times \mathrm{depth} \times \mathrm{length} = 0.8~\rm{m}\times0.6~\rm{m}\times4.0~\rm{m}$), with the flow speed kept at zero ($U_{\infty}=0$ m/s). A NACA 0012 wing, made of clear acrylic, is mounted vertically in the tunnel, with an endplate on the top to skim surface waves and eliminate wingtip vortices at the root. The wing is connected to a six-axis force/torque transducer (ATI 9105-TIF-Delta-IP65), which measures the fluid torque $\tau_f$ exerted on the wing. This $\tau_f$ is then fed into the cyber-physical system (CPS). Depending on the input virtual structural parameters, specifically the torsional stiffness $k_v$, damping $b_v$ and inertia $I_v$, the CPS calculates the pitching position of the wing and outputs the signal to the servo motor (Parker SM233AE). An optical encoder (US Digital E3-2500) which is independent of the CPS is used to measure the pitching position $\theta$. The CPS is operated at 4000 Hz to minimise any phase delay between the input $\tau_f$ and the output $\theta$. A detailed explanation of the CPS can be found in \citet{zhu2020nonlinear}.

We use two-dimensional particle image velocimetry (PIV) to measure the flow field around the wing. The flow is seeded using 50 $\mathrm{\mu m}$ diameter hollow ceramic spheres and illuminated by a laser sheet at the mid-span plane. The laser sheet is generated by a double-pulse Nd:YAG laser (532 nm, Quantel EverGreen) with LaVision sheet optics. The transparent wing enables flow field measurements on both sides of the wing. Due to the limitation of space beneath the tunnel, a $45^\circ$ mirror is used to reflect the images into two co-planar sCMOS cameras (LaVision). We use the DaVis software (LaVision) to calculate (two passes at $64\times64$ pixels, two passes at $32\times32$ pixels, both with 50\% overlap) and stitch the velocity fields from the two cameras to form a field of view of $3.2c \times 3.2c$, where $c$ is the chord length of the wing.

Figure \ref{fig.setup}(\emph{b}) sketches the two types of wings we use in the present study. For the unswept wing, a wing holder mechanism (not shown) enables the pivot axis to be adjusted between $x/c=0$ and 1 with a step size of 0.125. For the swept wings, the sweep angle $\Lambda$ is defined as the angle between the leading edge and the vertical axis. Four swept wings with $\Lambda = 10^\circ$, $15^\circ$, $20^\circ$ and $25^\circ$ are used. As shown in the figure, the pivot axis of swept wings is a vertical line passing through the mid-chord point ($x/c=0.5$) of the mid-span plane. All the wings have a span of $s=0.3$ m and a chord length of $c=0.1$ m, which results in an aspect ratio of $AR=3$.

The governing equation of the system is
\begin{equation}
	I \ddot{\theta} + b \dot{\theta} + k \theta = \tau_f,
	\label{eqn.govern}
\end{equation}
where $\theta$, $\dot{\theta}$, and $\ddot{\theta}$ are the angular position, velocity and acceleration, respectively. $I$, $b$ and $k$ are the effective inertia, damping and stiffness of the system. The effective inertia $I$ is the sum of the virtual inertia $I_v$, which we prescribe with the CPS, and the physical inertia $I_p$ of the wing (i.e. $I=I_v+I_p$). The effective damping $b$ equals the virtual damping $b_v$ (i.e. $b=b_v$) because the friction in the system is negligible. The effective stiffness $k$ equals the virtual stiffness (i.e. $k=k_v$). $\tau_f$ is the nonlinear fluid torque experienced by the wing, which can be divided into the added mass torque, $\tau_a = -I_a \ddot{\theta}$, where $I_a$ is the added fluid inertia, and the fluid damping torque, for simplicity $\tau_b = -b_f \dot{\theta}$, where $b_f$ is the fluid damping coefficient (see \S \ref{sec.intro}). Note that $b_f$ is expected to be a function of $\dot{\theta}$ \citep{mathai2019dynamics}. Equation \ref{eqn.govern} can thus be rearranged as
\begin{equation}
	(I + I_a) \ddot{\theta} + (b + b_f) \dot{\theta} + k \theta = 0.
	\label{eqn.govern_rearrange}
\end{equation}
After a perturbation of amplitude $A_0$ is applied at time $t_0$, the damped oscillations of the system can be described as
\begin{equation}
	\theta = A_0 e^{-\gamma (t-t_0)} \cos{[2\pi f_p (t-t_0)]},
	\label{eqn.damped_oscillation}
\end{equation}
where
\refstepcounter{equation}
$$
\gamma = \frac{b + b_f}{2(I + I_a)}~~\mathrm{and}~~
f_p = \frac{1}{2\pi}\sqrt{\frac{k}{I + I_a} - \gamma^2}.
\eqno{(\theequation{\mathit{a},\mathit{b}})}\label{eqn.fp}
$$

\section{Results and Discussion}
\subsection{Extracting the fluid damping from `ring down' experiments}{\label{sec.ringdown}}

\begin{figure}
\centering
\includegraphics[width=1\textwidth]{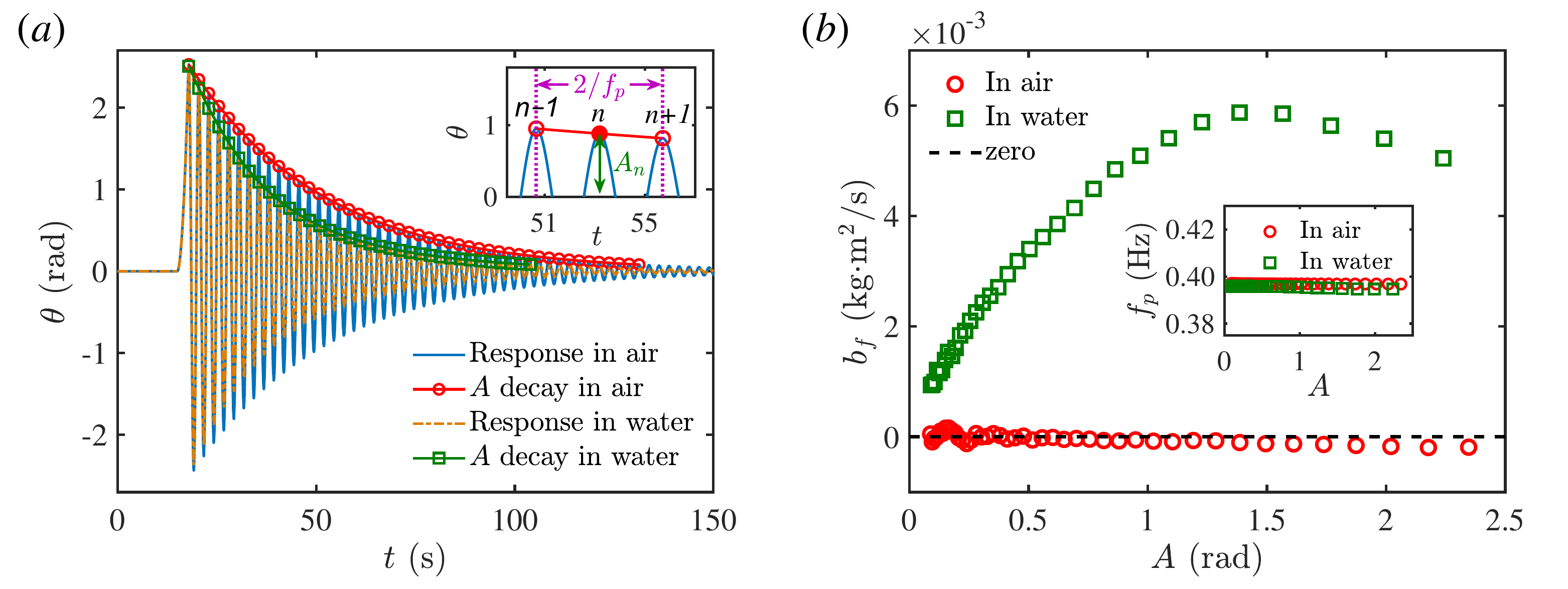}
\caption{(\emph{a}) System response and amplitude decay in a typical `ring down' test, where an elastically mounted unswept wing ($\Lambda=0^\circ$) pivots around the mid-chord ($x/c=0.5$) at a frequency of $f_p=0.40$ Hz. The inset shows the measurements of the pitching amplitude $A_n$ and the pitching frequency $f_p$ of the $n$-th peak. The fluid damping $b_f$ at $A_n$ is extracted by fitting an exponential curve (i.e. the red solid line) to the adjacent three peaks. (\emph{b}) Extracted $b_f$ in air and in water. The zero value is indicated by the black dashed line. The inset compares the measured pitching frequency $f_p$ in air and in water.}
\label{fig.extract_damp}
\end{figure}

We conduct `ring down' experiments to measure the fluid damping experienced by elastically mounted pitching wings. In the `ring down' experiment, a short-time constant-torque impulse is applied to the CPS as the perturbation, after which the system response and the amplitude decay of the wing is recorded and analysed. Figure \ref{fig.extract_damp}(\emph{a}) shows the results from a typical `ring down' experiment. In this specific case, we use an unswept wing ($\Lambda=0^\circ$) which pivots around the mid-chord ($x/c=0.5$) at a frequency of $f_p=0.40$ Hz. We conduct the `ring down' experiment twice -- once in air and once in water. The pitching amplitude of the wing decays faster in water than in air, indicating a higher total damping in water.

To quantify this amplitude decay, the positive peaks of the system response are identified. As shown in the inset, the amplitude of the $n$-th peak is denoted by $A_n$, and the corresponding pitching frequency is measured as $f_p = 2/(t_{n+1}-t_{n-1})$. To measure the total damping $b + b_f$ at amplitude $A_n$, we fit an exponential, $y=\alpha e^{-\gamma t}$, to the three adjacent peaks, $n-1$, $n$ and $n+1$, and extract the corresponding $\gamma$ (see equation \ref{eqn.damped_oscillation}). Now the only unknown in equation \ref{eqn.fp}\emph{b} is the added mass, $I_a$. After obtaining $I_a$, the fluid damping, $b_f$, is then calculated using equation \ref{eqn.fp}\emph{a} \citep{rao1995mechanical}. Since $f_p$ and $\gamma$ are both measured, $I_a$ and $b_f$ are also \emph{measured} quantities. Moreover, both $I_a$ and $b_f$ are \emph{cycle-averaged}, meaning they cannot reflect the instantaneous variation of the fluid inertia and damping. The measured fluid damping, $b_f$, in both air and water are compared in figure \ref{fig.extract_damp}(\emph{b}). Since $\tau_f$ in equation \ref{eqn.govern} is negligible in air as compared to other forces in the equation, $b_f$ stays near zero, which is indicated by the good agreement between the red circles and the black dashed line. As shown by the green squares, $b_f$ in water is significant because of the existence of the fluid damping torque, $\tau_{b}$. It is also observed that $b_f$ in water increases non-monotonically with $A$. This nonlinear behaviour will be revisited later in \S \ref{sec.flow_dynamics}. The inset of figure \ref{fig.extract_damp}(\emph{b}) shows the measured pitching frequency, $f_p$, in both air and water. Due to the combined effect of the fluid inertia and damping, we see that $f_p$ is slightly lower in water than in air.

\subsection{Frequency scaling of the fluid damping}{\label{sec.freq}}

\begin{figure}
\centering
\includegraphics[width=1\textwidth]{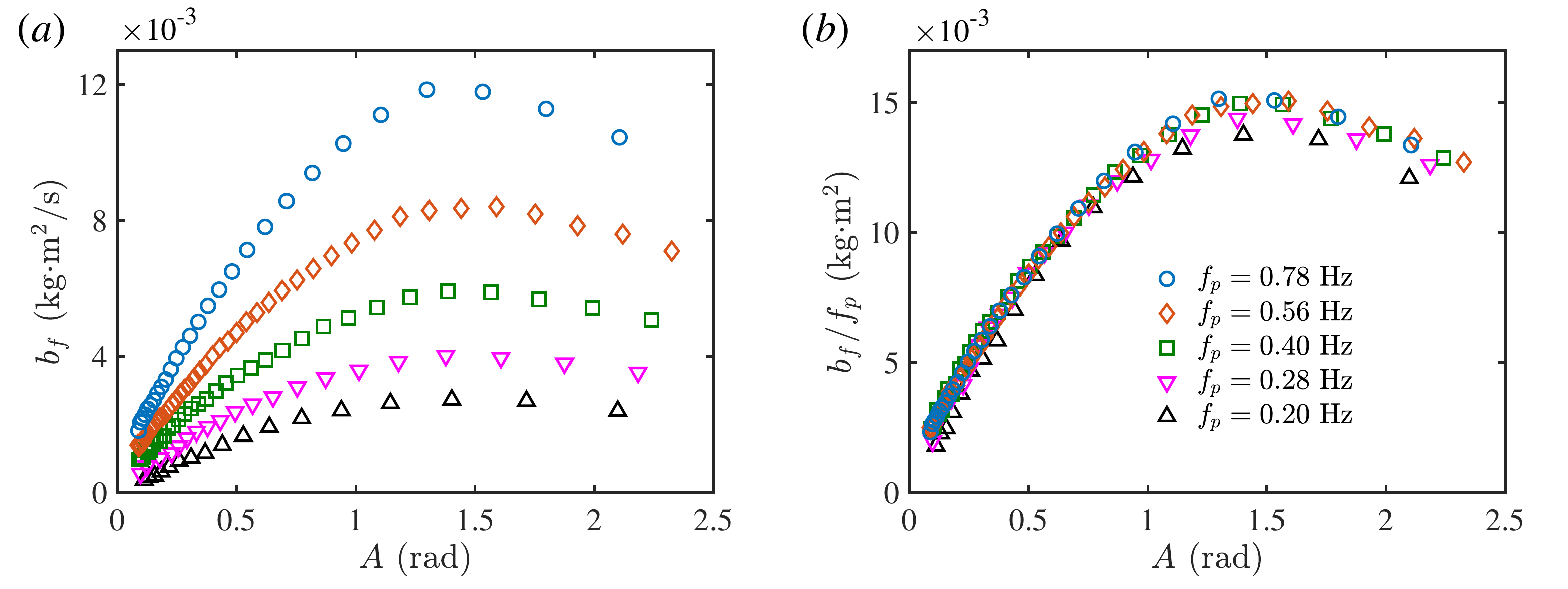}
\caption{(\emph{a}) Extracted fluid damping $b_f$ at different pitching frequencies for $x/c=0.5$. (\emph{b}) A frequency scaling for the fluid damping which collapses $b_f$ at different $f_p$ into one curve. Note that (\emph{a}) and (\emph{b}) share the same legend.}
\label{fig.damp_freq}
\end{figure}

We repeat the `ring down' experiment for the unswept wing ($\Lambda=0^\circ$) pivoting at the mid-chord, $x/c=0.5$, and change the pitching frequency by tuning the virtual inertia, $I_v$, and the virtual stiffness, $k_v$, while keeping the virtual damping $b_v$ constant \citep{onoue2016vortex,onoue2017scaling}. The extracted fluid damping, $b_f$, are shown in figure \ref{fig.damp_freq}(\emph{a}). Note that figure \ref{fig.damp_freq}(\emph{a}) and (\emph{b}) share the same legend. We observe that $b_f$ increases monotonically with the pitching frequency, $f_p$, and that the trend of $b_f$ remains consistent for all frequencies. This observation agrees with those observed in heaving rigid plates \citep{keulegan1958forces,shih1971drag}, where the fluid damping coefficient scales inversely with the oscillation period. As we discussed earlier, $b_f$ derives from the fluid damping torque $\tau_{b}$, which depends strongly on the vortex-induced forces on the wing \citep{kang2014analytical}. \citet{onoue2016vortex,onoue2017scaling} have shown that the circulation of LEVs scales with the strength of the feeding shear-layer velocity. In our case without a free-stream flow, the feeding shear-layer velocity equals the leading-/trailing-edge velocity, which is proportional to $f_p$. Based on this, we divide $b_f$ by $f_p$ (figure \ref{fig.damp_freq}\emph{b}). It is seen that with this scaling, all of the fluid damping curves collapse nicely.

\begin{figure}
\centering
\includegraphics[width=1\textwidth]{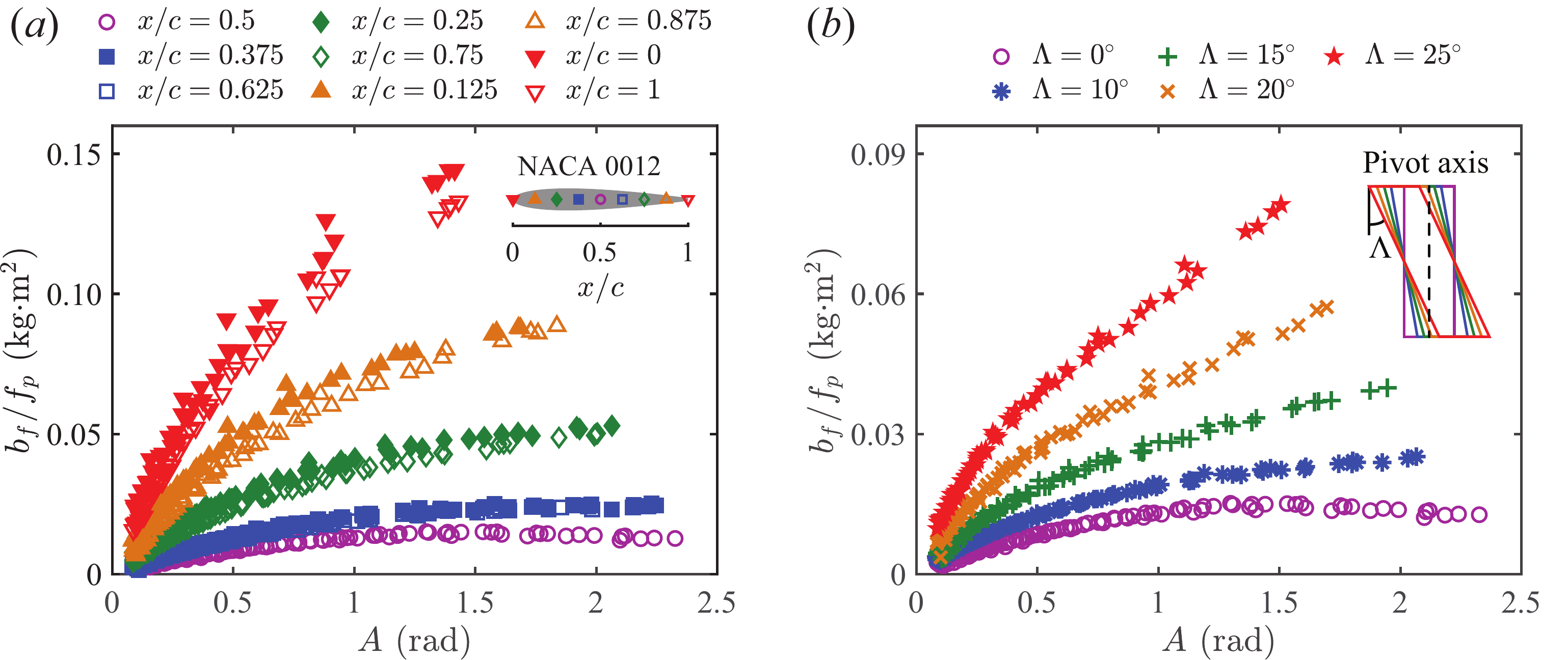}
\caption{(\emph{a}) $b_f/f_p$ for an unswept wing ($\Lambda=0^\circ$) pivoting at $x/c=0$ to 1 with a step size of 0.125. The pivot location for each dataset is shown by the inset. (\emph{b}) $b_f/f_p$ for swept wings with $\Lambda=0^\circ$, $10^\circ$, $15^\circ$, $20^\circ$ and $25^\circ$. The inset shows side views of the five swept wings and the dashed line indicates the pivot axis. The colours of the wings correspond to the colours of $b_f/f_p$ curves in the figure. The purple circles in (\emph{a}) and (\emph{b}) are replotted from figure \ref{fig.damp_freq}(\emph{b}). Note that each dataset in (\emph{a}) and (\emph{b}) includes \emph{five} different $f_p$.}
\label{fig.damp_axis_sweep}
\end{figure}

We extend this frequency scaling to unswept wings with different pivot axes (figure \ref{fig.damp_axis_sweep}\emph{a}) and to swept wings with different sweep angles (figure \ref{fig.damp_axis_sweep}\emph{b}). For comparison, we include the previous results (figure \ref{fig.damp_freq}\emph{b}) using purple circles in both figure \ref{fig.damp_axis_sweep}(\emph{a}) and (\emph{b}). Note that each symbol shape in figure \ref{fig.damp_axis_sweep} contains \emph{five} different pitching frequencies, $f_p = 0.20$, 0.28, 0.40, 0.56 and 0.78 Hz.

For the unswept wing ($\Lambda=0^\circ$), we change the pivot axis from $x/c=0$ to 1 with a step size of 0.125 (see the inset of figure \ref{fig.damp_axis_sweep}\emph{a}). We observe that $b_f/f_p$ increases as the pivot axis is moved away from the mid-chord, $x/c=0.5$. For pivot axes that are symmetric with respect to the mid-chord (i.e. $x/c=0.375$ \& 0.625, 0.25 \& 0.75, 0.125 \& 0.875 and 0 \& 1), $b_f/f_p$ roughly overlap. The slight inconsistency between $b_f/f_p$ for $x/c>0.5$ and $x/c<0.5$ comes from the asymmetry of the NACA 0012 wing geometry with respect to the mid-chord; we see that the scaled damping, $b_f/f_p$, is always slightly higher for $x/c<0.5$. In these cases, the damping at the trailing edge dominates due to the higher velocity and longer moment arm, and is stronger than the cases when $x/c>0.5$, where the leading-edge damping dominates. We will show in \S \ref{sec.flow_dynamics} that this is due to differences in the vortex structures generated by the sharp and rounded geometries.

This frequency scaling, $b_f/f_p$, also holds for three-dimensional (3D) swept wings (figure \ref{fig.damp_axis_sweep}\emph{b}). Again, each curve includes data from five pitching frequencies. Here, the pivot axes of swept wings are kept as a vertical line passing through the mid-chord of the mid-span plane (see the inset of figure \ref{fig.damp_axis_sweep}\emph{b}). As $\Lambda$ increases, the average pivot axes of the top and the bottom portion of the swept wing move away from the mid-chord, leading to the increase of the scaled damping, $b_f/f_p$, in a manner similar to that observed for unswept wings with different pivot locations (figure \ref{fig.damp_axis_sweep}\emph{a}). This argument will be revisited in the next section.

\subsection{Universal fluid damping scaling for unswept and swept wings}{\label{sec.scaling}}

\begin{figure}
\centering
\includegraphics[width=0.7\textwidth]{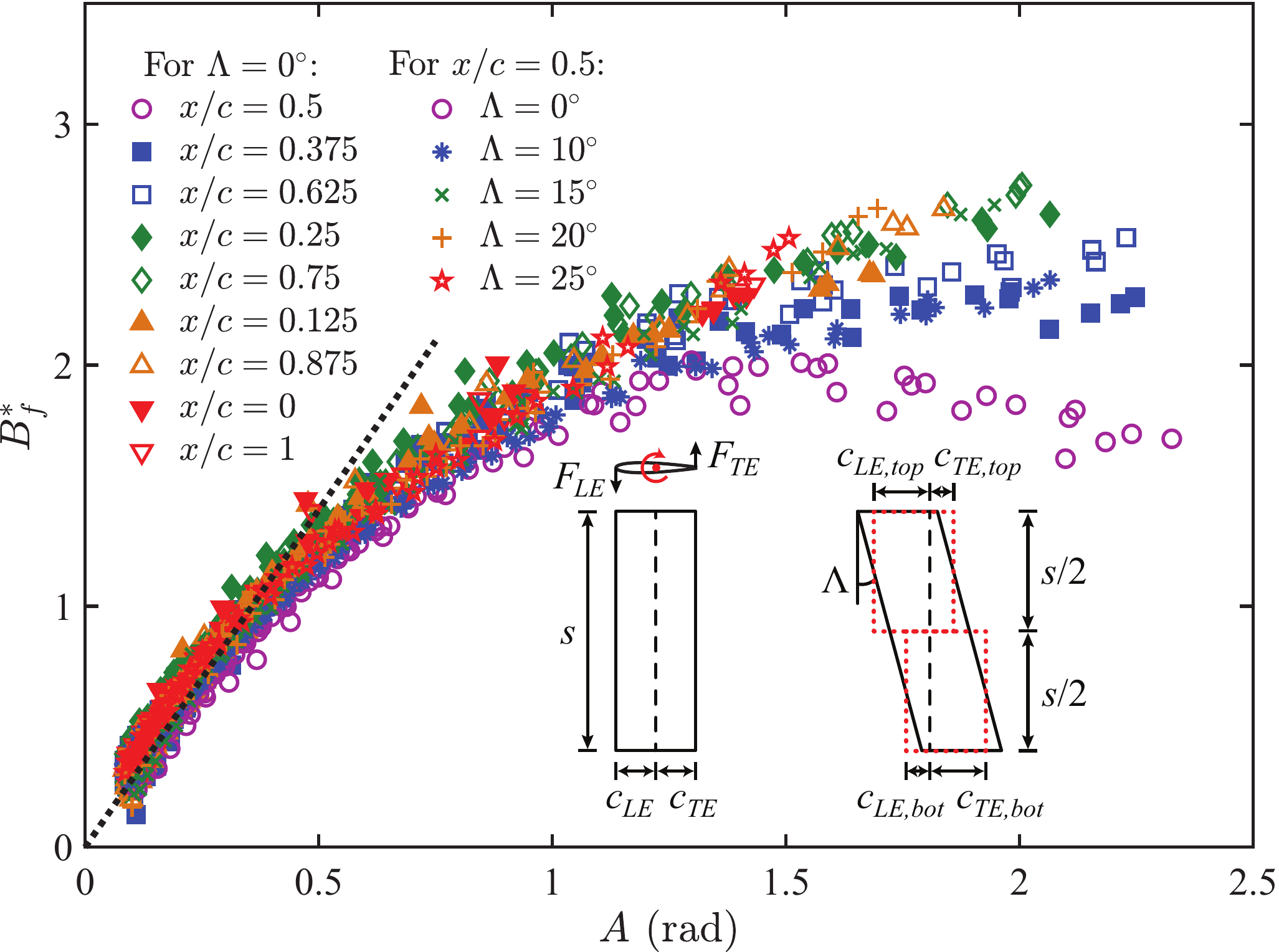}
\caption{Non-dimensional fluid damping coefficient $B_f^*$ versus pitching amplitude $A$ for unswept wings pivoting at $x/c=0$ to 1 and swept wings with sweep angles $\Lambda=0^\circ$ to $25^\circ$. The inset shows the definition of the leading-edge chord $c_{LE}$ and the trailing-edge chord $c_{TE}$, with black dashed lines indicating the pivot axes. The black dotted line indicates the small amplitude prediction for a drag coefficient of $C_D=2.8$.}
\label{fig.universal_scaling}
\end{figure}

Figure \ref{fig.damp_axis_sweep}(\emph{a}) indicates that the pivot axis plays an important role in determining the fluid damping of unswept wings. We extend the frequency scaling of $b_f$ to take into account this effect. First, we divide the wing into two parts, the fore part from LE to the pivot axis with a chord length of $c_{LE}$, and the aft part from the pivot axis to TE with a chord length of $c_{TE}$ (see the inset of figure \ref{fig.universal_scaling} for an example when the wing pivots at $x/c=0.5$). The Morison equation \citep{morison1950force} indicates that the fluid damping force $F$ scales with $0.5 \rho U^2 sc$, where $\rho$ is the fluid density, $U \sim \dot{\theta} c$ is the characteristic velocity and $sc$ is the wing area. We can express the total fluid damping torque as the sum of the torque exerted on the fore and aft portions of the wing,
\begin{equation}
	\tau_b \sim K_{LE} F_{LE} c_{LE} + K_{TE} F_{TE} c_{TE},
	\label{eqn.damping_torque}
\end{equation}
where the subscripts $LE$ and $TE$ refer to the leading- and trailing-edge contributions, and $K_{LE}$ and $K_{TE}$ are empirical factors that account for the subtle differences in the damping associated with the specific geometries of the leading and trailing edges (figure \ref{fig.damp_axis_sweep}\emph{a}). Since the differences are small, $K_{LE}$ and $K_{TE}$ should be close to one, and for consistency, their average value must equal one ($(K_{LE} + K_{TE})/2=1$).

Since the damped oscillations are observed to be near-sinusoidal (figure \ref{fig.extract_damp}\emph{a}), the average angular velocity is given by $4 f_p A$. Simplifying, we arrive at an expression for the fluid damping:
\begin{equation}
	b_f \sim 2 \rho f_p A s (K_{LE} c_{LE}^4 + K_{TE} c_{TE}^4),
	\label{eqn.b_f_scaling}
\end{equation}
or, in non-dimensional form,
\begin{equation}
	B_f^* \equiv \frac{b_f}{2 \rho f_p s (K_{LE} c_{LE}^4 + K_{TE} c_{TE}^4)} ~\propto A.
	\label{eqn.non-dimensional_B}
\end{equation}

For swept wings, because the pivot axis passes through $x/c=0.5$ at the mid-span, the top half of the wing has an average pivot axis $x/c>0.5$, while the bottom half has an average pivot axis $x/c<0.5$. Ignoring three-dimensional effects, we approximate the swept wing by two `equivalent' unswept wing segments. We choose not to divide the wing into a large number of narrow `blade elements' \citep{glauert1983elements}, because the pivot axis of some elements near the wing root/tip for large sweep angles may lie outside the range $x/c = [0,1]$, where our scaling has not been tested. The inset of figure \ref{fig.universal_scaling} shows how these two unswept wing segments are configured (rectangles with red dotted lines). Based on the wing geometry, we see that
\begin{equation}
	\begin{split}
		c_{LE,top} & = c_{TE,bot} = \frac{c}{2}+\frac{s}{4}\tan{\Lambda},\\
		c_{TE,top} & = c_{LE,bot} = \frac{c}{2}-\frac{s}{4}\tan{\Lambda}.
	\end{split}
	\label{eqn.swept_wing_chord}
\end{equation}
Following the same analysis as for the unswept wing, and adding the fluid damping of the top and the bottom wing segments together, we find that the fluid damping for the full swept wing is given by
\begin{equation}
	b_f \sim \rho f_p A s (K_{LE} c_{LE,top}^4 + K_{TE} c_{TE,top}^4 + K_{LE} c_{LE,bot}^4 + K_{TE} c_{TE,bot}^4).
	\label{eqn.b_f_scaling_swept}
\end{equation}
If we define an effective leading-edge chord $c_{LE} = c_{LE,top} = c_{TE,bot}$ and an effective trailing-edge chord $c_{TE} = c_{TE,top} = c_{LE,bot}$, this scaling reduces to equation \ref{eqn.b_f_scaling} with $K_{LE}$ and $K_{TE}$ cancelled out. This cancellation results because the effective pivot axes of the top and the bottom segments are symmetric about $x/c=0.5$ at the mid-span, which averages out the slight differences in fluid damping experienced by the top and the bottom segments. For the same reason, $K_{LE}$ and $K_{TE}$ also cancel out in equation \ref{eqn.non-dimensional_B} for swept wings.

Figure \ref{fig.universal_scaling} shows the non-dimensional fluid damping, $B_f^*$, as a function of the pitching amplitude, $A$, for unswept and swept wings. Here, we have used $K_{LE} = 0.95$ and $K_{TE} = 1.05$. We see that all of our measurements collapse remarkably well under the proposed scaling, especially for $A<1.57~(90^\circ)$, despite the wide range of pitching frequencies ($f_p = 0.20$ to 0.78 Hz), pivot axes ($x/c=0$ to 1) and sweep angles ($\Lambda = 0^\circ$ to $25^\circ$) tested in the experiments. In the small-amplitude limit ($A<0.5$), $B_f^*$ scales linearly with $A$, with a slope that corresponds to the drag coefficient, $C_D$. We note that $C_D \approx 2.8$, which is comparable to that of an accelerated normal flat plate \citep{ringuette2007role}. At higher pitching angles ($A>0.5$), however, the linear approximation no longer holds and we see a decreasing slope of $B_f^*$ as a function of $A$. This is presumably because the shed vortices no longer follow the rotating wing and the fluid force becomes non-perpendicular to the wing surface as $A$ increases. For $A>1.57~(90^\circ)$, the scaling works reasonably well except for the case $\Lambda=0^\circ$, $x/c=0.5$, where a decreasing $B_f^*$ is observed. In the next section, we will use insights from the velocity fields to explain this non-monotonic behaviour.

\subsection{Insights obtained from velocity fields}{\label{sec.flow_dynamics}}

\begin{figure}
\centering
\includegraphics[width=0.9\textwidth]{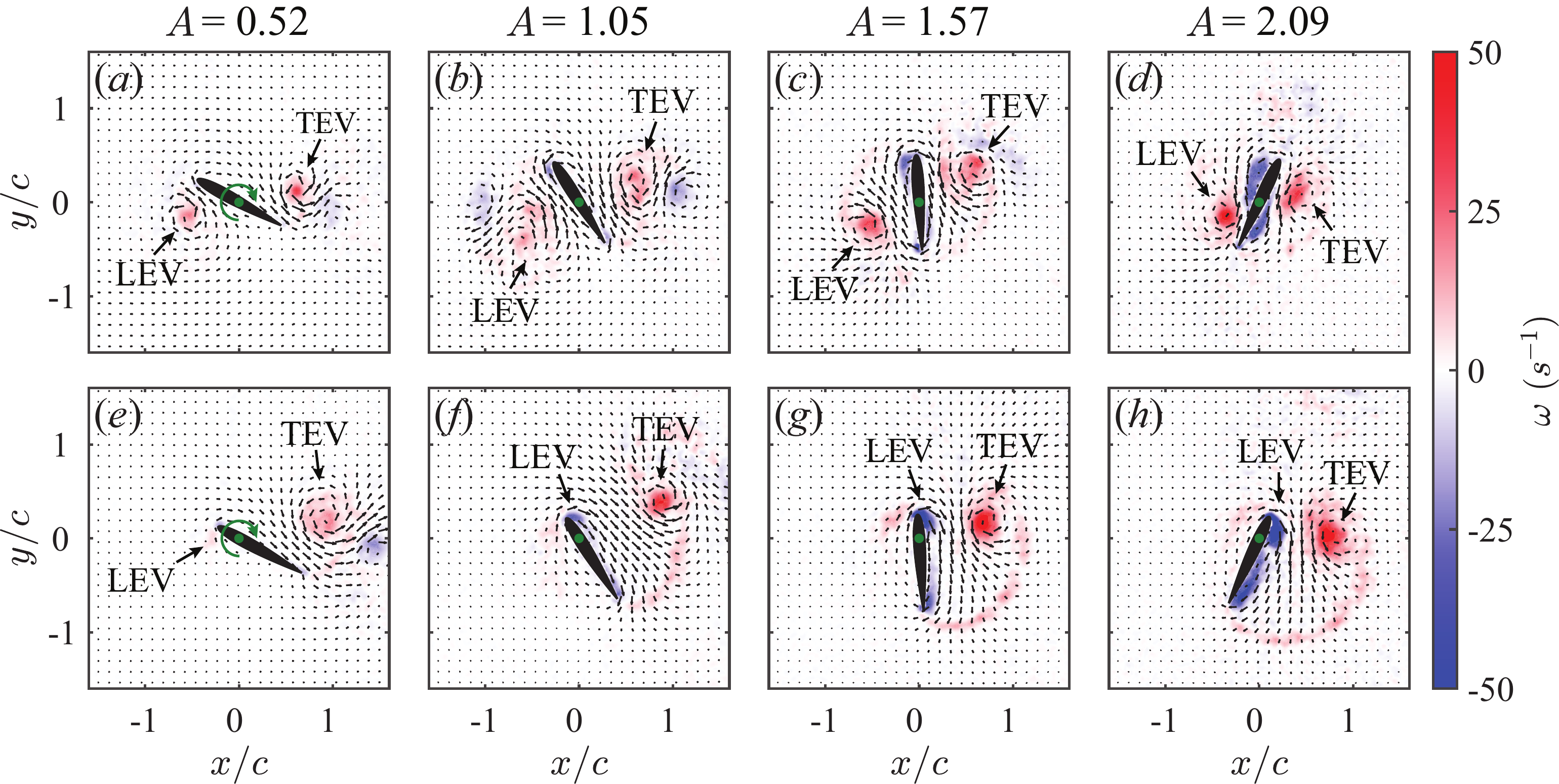}
\caption{PIV flow field measurements for an unswept wing undergoing prescribed sinusoidal pitching motions in quiescent water. (\emph{a--d}) Pivot axis (shown by green dots) $x/c=0.5$, pitching frequency $f_p=0.5$ Hz, pitching amplitude $A=0.52~(30^\circ),~1.05~(60^\circ),~1.57~(90^\circ)$ and $2.09~(120^\circ)$. (\emph{e--h}) Same as (\emph{a--d}), except that the pivot axis is at $x/c=0.25$. All the velocity fields are phase-averaged over 20 cycles. Only every fifth velocity vector is shown. Spanwise vorticity $\omega$: positive (red), counterclockwise; negative (blue), clockwise. See supplementary materials for the full video.}
\label{fig.piv}
\end{figure}

To gain more insight regarding the nonlinear behaviour of $B_f^*$, we conduct 2D PIV experiments to measure the surrounding flow fields of an unswept wing ($\Lambda=0^\circ$) with a prescribed pitching motion: $\theta = A \sin{(2 \pi f_p t)}$. The results are shown in figure \ref{fig.piv}. The pitching frequency is kept at $f_p=0.5$ Hz for all the cases and the pitching amplitude is varied from $A=0.52~(30^\circ)$ to $2.09~(120^\circ)$ with a step size of $0.52~(30^\circ)$. Two pivot axes are tested, $x/c=0.5$ (figure \ref{fig.piv}\emph{a--d}) and $x/c=0.25$ (figure \ref{fig.piv}\emph{e--h}). Note that the flow fields shown in figure \ref{fig.piv} are \emph{not} sequential. Instead, all the snapshots are taken right before $t/T=0.25$ for different pitching amplitudes, where $T$ is the pitching period. This specific time instant is chosen because it best reflects the difference in dynamics associated with the different pitching amplitudes and pivot axes.

For both pivot locations (figure \ref{fig.piv}\emph{a--d}: $x/c=0.5$ and \emph{e--h}: $x/c=0.25$), the spanwise vorticity of the pitch-generated leading-edge vortex (LEV) and trailing-edge vortex (TEV) increases with the pitching amplitude, $A$. This can be explained by the increase in the feeding shear-layer velocities associated with the higher pitching amplitudes \citep{onoue2016vortex}. The boundary vortices near the wing surface, which are related to the added mass effect \citep{corkery2019quantification}, also become more prominent due to the increase of the angular acceleration. When the wing pivots at $x/c=0.5$ (figure \ref{fig.piv}\emph{a--d}), the leading-edge velocity equals the trailing-edge velocity. As a result, the LEV and TEV are fairly symmetric about the pivot axis, with some subtle differences caused by the rounded and sharp edges, respectively. This confirms the arguments given earlier for the differences between $b_f/f_p$ for $x/c>0.5$ and $x/c<0.5$ (figure \ref{fig.damp_axis_sweep}\emph{a}). For $x/c=0.25$ (figure \ref{fig.piv}\emph{e--h}), however, the TEV is much more prominent than the LEV because of the higher trailing-edge velocity. Due to the low leading-edge velocity and the pitch-induced rotational flow, the sign of the LEV even reverses and becomes negative for $A=1.05$ to $2.09$ (figure \ref{fig.piv}\emph{f--h}).

For both pivot locations, due to the absence of a convective free stream, and the existence of the pitch-induced rotational flow, the LEV and TEV (only the TEV for $x/c=0.25$) are entrained closer to the wing surface as $A$ increases. For $x/c=0.5$, as shown in figure \ref{fig.piv}(\emph{c--d}), the LEV moves towards the aft portion of the wing and the TEV moves towards the fore portion of the wing when $A \geq 1.57~(90^\circ)$. The torque generated by these two vortices, which counteracts the wing rotation for small $A$, now assists the rotation as the wing pitches up towards higher angular positions. This assist reduces the fluid drag experienced by the wing and thus lowers the fluid damping. This effect can account for the non-monotonic behaviour of $B_f^*$ for $x/c=0.5$ (figure \ref{fig.universal_scaling}). For $x/c=0.25$ (figure \ref{fig.piv}\emph{g--h}), a similar scenario is observed, in which the TEV moves towards the fore portion of the wing and gets closer to the wing surface as $A$ increases. However, because of the existence of a counter-rotating LEV, the TEV is not able to approach the wing surface as closely as in the $x/c=0.5$ case. This explains why a flattening behaviour, rather than a non-monotonic trend of $B_f^*$, is observed for $x/c=0.25$ and presumably for other pivot locations at high pitching amplitudes.

\section{Conclusions}\label{sec.conclusion}

By utilising a cyber-physical control system to create an elastically mounted pitching wing, we have experimentally measured the nonlinear fluid damping associated with vortices shed from a bluff body. A theoretical scaling has been proposed and validated, based on the Morison equation, which incorporates the frequency, amplitude, pivot location and sweep angle. The nonlinear behaviour of the scaled fluid damping has been correlated with the velocity fields measured using particle image velocimetry.

One should note that our scaling may not be applicable for instantaneous fluid damping, because the damping characterised in the present study is cycle-averaged over near-sinusoidal oscillations. In addition, we have not considered three-dimensional effects, which are present due to the wing tip flows. Incorporating these may further improve the collapse of the fluid damping coefficient, $B_f^*$ (figure \ref{fig.universal_scaling}). Lastly, in \S \ref{sec.flow_dynamics}, only qualitative analysis of the flow field has been performed thus far. In order to get more accurate correspondence between the fluid damping and the flow dynamics, quantitative analysis of the vortex trajectory and circulation is needed, which will be the focus of future study.

Despite these limitations, the proposed scaling has been shown to collapse the data over a wide range of operating conditions ($f_p=0.20$ to 0.78 Hz and $A=0$ to 2.5) for both unswept ($x/c=0$ to 1) and swept wings ($\Lambda=0^\circ$ to $25^\circ$). It can be used to predict damping associated with shed vortices, and thus benefit the future modelling of a wide variety of flows, including unswept and swept wings in unsteady flows as well as other bluff body geometries. The universality of this scaling reinforces the underlying connection between swept wings and unswept wings with different pivot locations. In addition, the results presented in this study will be of potential value as a source of experimental data for validation and comparison of future theoretical/computational models.

\section*{Acknowledgments}
 
This work is funded by Air Force Office of Scientific Research, Grant FA9550-18-1-0322, managed by Dr. Gregg Abate.


\end{document}